\documentclass[twocolumn,prl,showpacs,floatfix]{revtex4-1}
\usepackage{graphicx}
\usepackage{epsfig}
\usepackage{rotating}
\usepackage{amsmath}
\usepackage{amsfonts}
\usepackage{amssymb}
\usepackage{enumerate}
\usepackage{longtable}
\usepackage{dcolumn}
\usepackage{bm}
\usepackage{physics}
\usepackage{verbatim}
\newcommand{\zncu}{ZnCu$_3$(OH)$_6$Cl$_2$ }

\begin{document}

\title{Nuclear relaxation rates in the Herbertsmithite Kagome antiferromagnets \zncu}

\author{Nicholas E. Sherman,$^1$ Takashi Imai$^{2,3}$  and Rajiv R. P. Singh$^1$}
\affiliation{$^1$Department of Physics, University of California Davis, CA 95616, USA }

\affiliation{$^2$Department of Physics and Astronomy, McMaster University, Hamilton, Ontario L8S4M1, Canada }

\affiliation{$^3$Canadian Institute for Advanced Research, Toronto, Ontario M5G1Z8, Canada }

\date{\rm\today}

\begin{abstract}
Local spectral functions and Nuclear Magnetic Relaxation (NMR) rates, $1/T_1$, for the spin-half Heisenberg antiferromagnet on the Kagome Lattice 
are calculated using Moriya’s Gaussian approximation, as well as through an extrapolation of multiple frequency moments. The temperature dependence 
of the calculated rates is compared with the oxygen $1/T_1$ NMR data in Herbertsmithite. We find that the Gaussian approximation for
$1/T_1$ shows behavior qualitatively similar to experiments with a sharp drop in rates at low temperatures, consistent with a spin-gapped behavior.
However, this approximation significantly underestimates the magnitude of $1/T_1$ even at room temperature.
Rates obtained from extrapolation of multiple frequency moments give very good agreement with the room temperature NMR data with $J = 195 \pm 20 K$ and hyperfine couplings 
determined independently from other measurements. The use of multiple frequency moments also leads to additional low frequency weight in the local structure factors.
The convergence of our calculations with higher
frequency moments breaks down at low temperatures suggesting the existence of longer range dynamic correlations in the system despite
the very short-range static correlations.

\end{abstract}


\maketitle
{\it Introduction}:
The Kagome-lattice Heisenberg antiferromagnets remain one of the strongest experimental candidates for a quantum spin-liquid \cite{balents,imai-lee}.
Recent computational studies using Density Matrix Renormalization Group (DMRG)
have presented strong evidence for a $Z_2$ quantum spin-liquid ground state 
with a small spin-gap  of order $0.1 $ J \cite{dmrg}, supplanting previous theoretical support for
Valence Bond Solids, gapless spin-liquids and other candidate phases\cite{lhuillier,waldtmann,mila,vishwanath,elstner,vbc,ran07,singh-huse-qc},
although some debate about the existence of a spin-gap and the nature of the spin-liquid phase remains \cite{gapless,nakano}. 

On the experimental side, the Herbertsmithite materials \zncu present a structurally perfect Kagome spin-lattice \cite{klhm-e}.
In the absence of antisite disorder mixing zinc and copper atoms, the material consists of undistorted Kagome planes
of copper spins, which are separated by non-magnetic triangular planes containing only zinc transition-metal atoms, 
thus leading to magnetically well isolated two-dimensional Kagome antiferromagnets.
Various experimental probes have clearly established the absence of long-range magnetic order down to temperatures
several orders of magnitude below the exchange energy scale $J\approx 200 K$ \cite{klhm-e,Imai2008,rigol07,sindzingre07}. 
The presence of antisite disorder, primarily substituting
zinc atoms by copper atoms, leads to an excess of free spins and muddies the question of spin-gap and low frequency behavior of the system
crucial to a precise characterization of the phase of the material.
However, higher energy inelastic neutron scattering spectra on single crystals show
a wave-vector dependence that is nicely captured by a Schwinger-Boson based Z$_2$ quantum spin-liquid calculation \cite{neutron,sachdev}. 

While the static spin-spin correlation length in the Kagome antiferromagnets never grows much bigger than a lattice spacing, the 
dynamic correlations can be 
longer-ranged. In Herbertsmithite, some power-laws in temperature
and frequency were reported in early neutron scattering experiments \cite{scaling}, 
though it has become rather clear that they primarily come from defects \cite{klhm-e,kawamura,vbg}.

Nuclear Magnetic Resonance (NMR) experiments have the unique ability to sense local environments and, hence, cleanly separate the
intrinsic Kagome spin response from impurities between kagome and triangular planes. Recent
NMR measurements have shown a spin-gap in the intrinsic Kagome spin response, strengthening the case for
a gapped spin-liquid \cite{FuScience}. To our knowledge, there are no previous theoretical calculations of NMR rates
in the system.

Here, we study the spin spectral functions and nuclear relaxation rates by a systematic computational method.
We use Moriya's Gaussian approximation based on a short-time expansion, to calculate the Nuclear relaxation rates \cite{moriya,Imai1993,gelfand}.
These calculations, done 
using Numerical Linked Cluster (NLC) expansions \cite{nlc-et}, show good internal convergence down to low temperatures.
They show behavior that is qualitatively similar to the experimental data with a sharp drop towards vanishing rates
below a temperature of order $0.1 J$.
We then present calculations of the local spectral
functions relevant to NMR that go beyond the short-time expansion by calculating multiple frequency moments 
of the spectral function through
NLC. The spectral lineshapes are then deduced using suitable ansatz for the
frequency dependence. 
The resulting structure factors show additional low frequency weight.

The resluting nuclear relaxation rates can now be quantitatively compared with oxygen NMR measurements \cite{FuScience} in \zncu. For the
comparison, the hyperfine couplings are calculated from measurements of uniform susceptibility and Knight shift \cite{FuScience,Imai2011}, leaving only
the exchange constant $J$ as the free parameter. We find that there is
very good quantitative agreement between theory and experiments with $J=195\pm 20$ K  \cite{valenti}.
We also find that our calculation of $1/T_1$ based on multiple frequency moments loses convergence below a temperature of $0.2 J$ just as
the rates show evidence for sharp downturn with temperature. Thus we are unable to theoretically settle the existence of a spin-gap
in this system. The breakdown of NLC implies that the dynamic or higher energy spin correlations are becoming
longer-ranged in the model despite very short-ranged static correlations.

{\it Model and Spectral Functions}:
We consider the Heisenberg model with Hamiltonian:
\begin{equation}
{\cal H}= J\sum_{\langle i,j \rangle} (S_i^x S_j^x + S_i^y S_j^y+S_i^z S_j^z),
\end{equation}
where the sum runs over all nearest-neighbor bonds of the Kagome lattice.
The $S_i^\alpha$ ($\alpha=x,y,z$) represent spin-half operators associated with
the spin at site $i$. For theoretical calculations we set $J=1$.

The nuclear relaxation rate, $1/T_1$, for a nucleus is proportional to the Fourier transform of the
dynamic correlation function,
\begin{equation}
S_a(t)=\langle O_a(t)O_a(0)\rangle ,
\end{equation}
which we denote $\hat S_a(\omega)$ evaluated at the nuclear resonance frequency $\omega_N<<J$. 
The angular brackets refer to thermal averaging with respect
to the canonical distribution. The operators $O_a$ depend on the nucleus under consideration. 
We focus here on the oxygen nucleus for which accurate measurements of hyperfine couplings exist. The geometry
of Herbertsmithite dictates that the simplest choices for the operator for the oxygen nucleus is
\begin{equation}
O_{o}=S_i^z +S_j^z,
\end{equation}
where $i$ and $j$ are neighboring copper atoms equidistant from the oxygen nucleus. 
Spin rotational
invariance of the Heisenberg model means that it is sufficient to consider the operators as pointing along $z$ axis.

The Gaussian approximation is based on a short-time expansion of the correlation function (setting $\hbar=1$)
\begin{equation}
S_a(t) = c_0 + {-it \over 1 !}c_1 + {(-it)^2\over 2 !} c_2 + \ldots.
\end{equation}
The quantities $c_i$ can be shown to be the frequency moment of $\hat S_a(\omega)$. Writing a Gaussian with this short-time
expansion to order $t^2$, and defining 
\begin{equation}
m_2=c_2 c_0 - c_1^2,
\end{equation}
the zero frequency $\hat S_a(\omega)$, that is proportional to $1/T_1$, is given by
\begin{equation}
\hat S_a(0) = c_0^2\sqrt{\frac{2\pi}{m_2}}\exp(-\frac{c_1^2}{2m_2}).
\end{equation}

To study the frequency dependence in more detail, we define a quantity
that is even in $\omega$ so that its extrapolation in $\omega$ as an even function will maintain
the correct fluctuation-dissipation relation, but also has enhanced weight at low frequencies:
\begin{eqnarray}
 \hat K_a(\omega) &={1\over Z} \sum_{n,m} {\exp(-\beta E_m) -\exp(-\beta E_n)\over E_n -E_m} \\ \nonumber
                  &\times |\langle m|O_a|n\rangle |^2 \delta({\omega-E_n +E_m}),
\end{eqnarray}
where $Z$ is the partition function, $\ket{m}$ are the eigenstates of the Hamiltonian with eigenvalues $E_m$ 
and the case of $E_m\to E_n$ is treated in the limiting way as
\begin{equation}
 {\exp(-\beta E_m) -\exp(-\beta E_n)\over E_n -E_m} \longrightarrow \beta \exp(-\beta E_n). 
\end{equation}
The structure factors can be recovered from the spectral functions via the relation
\begin{equation}
\hat S_a(\omega) ={\omega \over 1 -\exp(-\beta\omega)} \hat K_a(\omega)
\end{equation}

Before proceeding with the numerical calculations, we note that our approach cannot capture the logarithmic divergence expected at low frequencies due to spin conservation and diffusion.
However, this divergence is rounded off due to anisotropies \cite{klhm-e}, and would further decrease as the temperature is lowered and
the spectral weight moves away from $q=0$. Even in the
one-dimensional Heisenberg model, where a much stronger square-root divergence is expected, once short range order builds up, the
diffusion peak becomes unnoticeable in computational approaches \cite{sandvik}.

{\it Numerical Linked cluster method:}
We use the Numerical Linked Cluster(NLC) method \cite{nlc-et} to calculate both the short time expansion coefficients for
the correlation functions $S_a(t)$ and the even order frequency moments $\mu_i$ for the spectral functions $\hat K_a(\omega)$.
Since these quantities can be expressed as thermal expectation values, they have a well-defined high temperature
expansion and a Numerical Linked Cluster expansion in the thermodynamic limit. 
The NLC method uses the graphical basis of high temperature
expansions, where the quantity of interest is expressed as a sum over all linked clusters
\begin{equation}
P({\cal L})=\sum_c L(c)\times W(c).
\end{equation}
The sum over $c$ runs over all distinct clusters of the lattice ${\cal L}$.
$L(c)$ is called the lattice constant of the cluster c, and is the number of embeddings of the cluster 
in the lattice per site (or per unit cell). The quantity $W(c)$ is called the weight of the cluster and
is determined entirely by a calculation of the property on the finite cluster $c$ and all its sub-clusters. 
It is defined as
\begin{equation}
W(c)=P(c)-\sum_s W(s),
\end{equation}
where the sum over $s$ is over all proper subclusters of the cluster $c$. In a high temperature expansion, the
property $P(c)$ is expanded in powers of inverse temperature. In NLC, one carries out a calculation at a given
temperature by a numerically accurate treatment of the finite cluster.

For the Kagome lattice, it is useful to consider clusters made up of complete triangles only. It was found in Ref.~\onlinecite{nlc-et}
that whereas an NLC based on bond or site based graphs starts to break down as soon as the high temperature expansion
diverges, the triangle based NLC converges down to much lower temperatures. 
Calculations are done up to $8$ triangles or $8$th order, needing $34$ topologically distinct graphs.

As one goes to higher moments, their expression in terms of thermal expectation value involves 
more and more spin operators. Thus, their convergence within NLC is likely to
break down earlier. We found that, in $8$th order, up to 8th moment are well converged at least down to $T= 0.2 J$ (See Fig.~1).

\begin{figure}
\begin{center}
\includegraphics[width=0.7\linewidth]{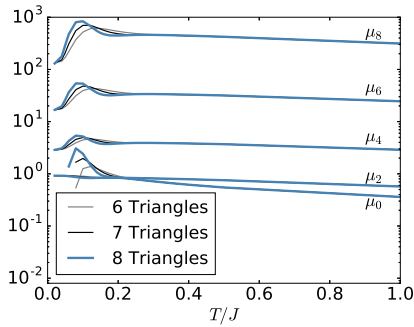} 
\caption{\label{moments} Frequency moments of the spectral function $\hat{K_o}$ for Oxygen NMR. The 6th, 7th and 8th order calculations are shown. Note that they
converge quite well down to a temperature of $0.2$ J.
}
\end{center}
\end{figure}

Reconstructing a distribution from its moments is, in general, an ill-posed problem . One scheme is the method of maximum entropy \cite{numerical-recipes}, but this method did not prove to be most useful to us. Instead, we assume a trial functional form $\hat{K_a^t}$ for $\hat K_a $, with tunable parameters $a_1,\cdots,a_n$ with $n$ the number of computed moments.  We then compute trial moments $\mu_i^t$ of $\hat{K_a^t}$.  This converts the problem of constructing the best $\hat{K_a^t}$ into a minimization problem, where we find the $\{a_i\}$ that minimizes our cost function $C(\{a_i\})$ defined by
\begin{equation}
C(\{a_i\}) = \sum_{i}\left(\frac{\mu_i-\mu_i^t}{\mu_i}\right)^2.
\end{equation}
Note that, if the fit works well, the minimum of such a function occurs when $\mu_i = \mu_i^t$ for each such $i$, and the value of $C$ is 0.  

We discuss two trial functions. First a triple Gaussian $\hat{K_{a}^G}$, with one Gaussian centered at $\omega=0$ and the other two symmetrically at a positive and a negative $\omega$ value.
This trial function gave NMR results close to Moriya's Gaussian approximation. However,
the cost function  $C$ was only on the order of 0.01,  implying that it did not reproduce all the moments
accurately. 

The second is a Lorentzian function of the form
\begin{align}
\hat{K_{a}^L} &= \frac{a_1}{a_2+\omega^2} 
\sum_{s\in \{-1,1\}}\exp\boldsymbol{\left(}-\left( \frac{\omega + sa_3}{a_4} \right)^{a_5}\boldsymbol{\right)}.
\end{align}
Since we know that all frequency moments are finite, the Lorentzian line-shape was cutoff in a suitable manner.
$\hat{K_{a}^L}$ had remarkable numerical success, giving values of the cost function $C$ on the order of $10^{-10}$. 

\begin{figure}
\begin{center}
\includegraphics[width=0.49\linewidth]{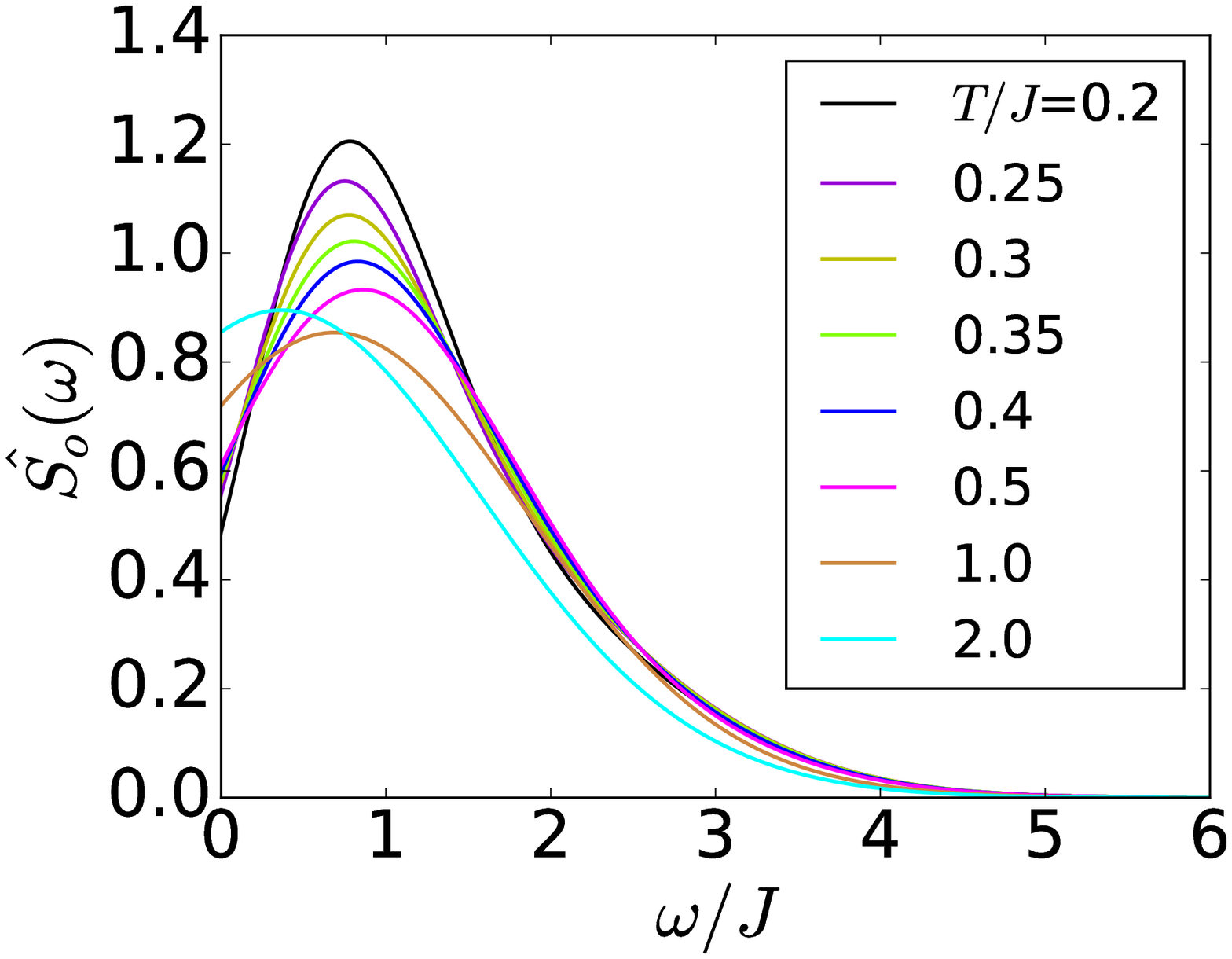} 
\includegraphics[width=0.49\linewidth]{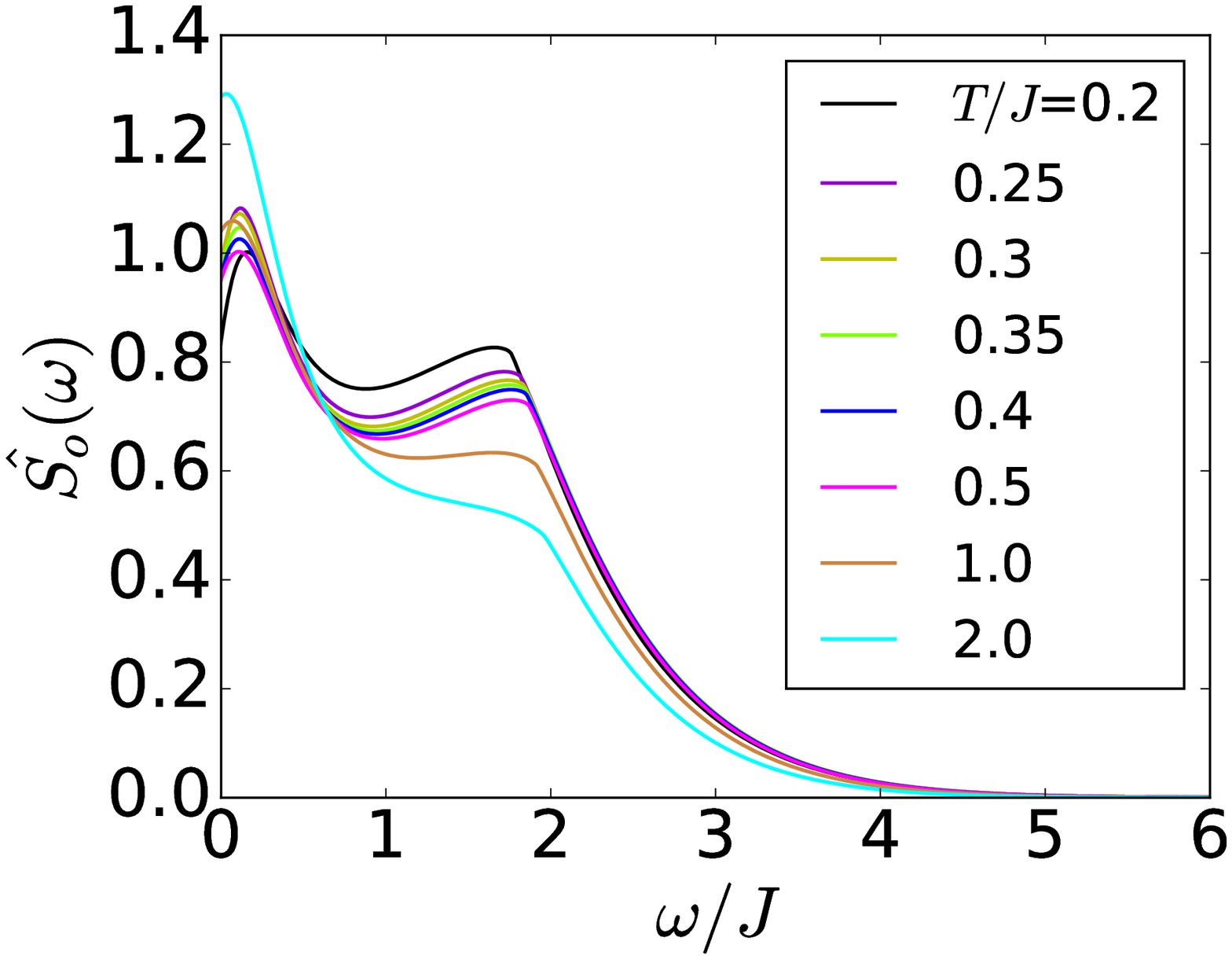} 
\caption{\label{TK} Plots of the frequency dependence of the bond structure factor relevant to oxygen NMR.
On the left we have results for $\hat S_o$ from the Gaussian spectral function $\hat{K_{o}^G}$ which is dominated by short time behavior. On the right we have results 
for $\hat S_o$ from the Lorentzian spectral function $\hat{K_{o}^L}$, which
gives a much better fit to the frequency moments.  
}
\end{center}
\end{figure}

We show structure factors obtained from both reconstructions of $\hat{K_o}$ in Fig.~\ref{TK}.  
In our fit for the Lorentzian lineshape, the value of $a_3$, for all temperatures
studied, was roughly 2.  This implies a rapid cutoff of spectral weight for $\omega/J > 2$.
If we consider a simple Ising model on the kagome lattice, when one spin is flipped, it would change energies of at most 4 bonds, 
corresponding to a maximum energy difference of $2J$.  The Heisenberg model has more complex behavior as a spin-flip can couple the many-body
states with arbitrary different energies. Nevertheless, we find that the spectral weights die off rapidly for frequencies larger than $2J$ in 
our model. 

The decay exponent $a_5$ was mostly around 1.4 in our fits and always between 1 and 2
as found in spin chains \cite{sandvik}. Both plots show a peak at frequencies away from zero, but the Lorentzian lineshape shows an additional peak at 
low frequencies. This low frequency peak is missed by the Gaussian lineshape and Moriya's Gaussian approximation.
This peak may gradually move away from zero as temperature is decreased, in a spin gap system.

\begin{figure}
\begin{center}
\includegraphics[width=0.7\linewidth]{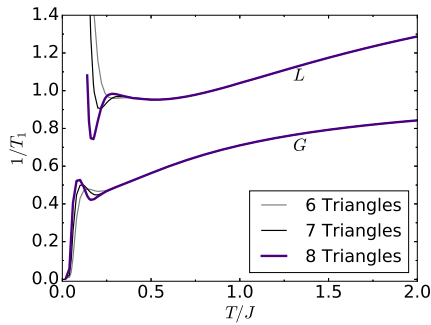} 
\caption{\label{LG} $1/T_1$ for the oxygen nucleus from Moriya's Gaussian approximation and the Lorentzian lineshape obtained from fitting all the frequency moments.
}
\end{center}
\end{figure}

A comparison of the NMR rates calculated by the Lorentzian lineshape and Moriya's Gaussian approximation is shown in Fig.~3. We see that while the temperature dependence
is similar in the two cases, the Gaussian approximation consistently underestimates the rates by up to a factor of 2. One advantage of the Gaussian approximation is
that, since it is based on the short time behavior, it can be continued all the way to $T=0$, whereas the more detailed extrapolation clearly breaks down at lower temperatures.



{\it Comparison with experiments}:
To compare our results with the experimental data we need the hyperfine couplings. For oxygen, the hyperfine Hamiltonian can be written as
\begin{equation}
{H}_{hf} =\sum_{m=1,2}\sum_{\alpha=i,j,k}B_{(\alpha)}I^{\alpha}S_{m}^{\alpha},
\end{equation}
where $B_{(\alpha)}$ represents the hyperfine coupling through the electron spin polarization induced in the 2s and/or 2p orbitals of the $^{17}$O sites.  The sum over
$m=$ $1$, $2$ refers to the two equidistant copper spins from the oxygen nucleus. Following the procedure outlined in \cite{gelfand}, the high temperature limit of $1/T_{1}$, in Moriya's Gaussian approximation, is given by
\begin{equation}
\frac{1}{T_{1M}}(T=\infty) = \frac{\sqrt{\pi}\{B_{(j)}^{2}+B_{(k)}^{2}\}}{8 \hbar J}\frac{4}{\sqrt{3}}.
\end{equation}
This sets the overall normalization for the rates in terms of the hyperfine couplings and exchange constant $J$.

\begin{figure}
\begin{center}
\includegraphics[width=0.7\linewidth]{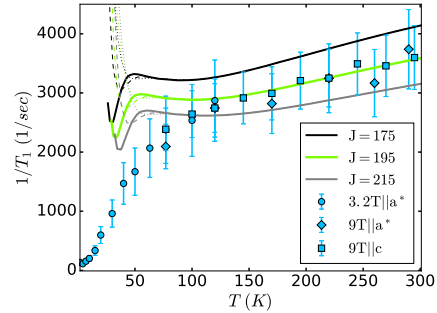} 
\caption{\label{compare} $1/T_1$ for the oxygen nucleus as a function of temperature 
computed using the Lorentzian lineshape and our calculated frequency moments compared with the experimental data on Herbertsmithite. 
The solid lines are the 8th order, dashed lines 7th order and dotted lines 6th order calculations. The lower orders can only be seen when they differ from the 8th order.
The experimental data shows some dependence on the orientation of the magnetic field (for details see \cite{FuScience} and supplementary materials), we rescaled the data for the field along $a$ to have the same value as the field along c at $T=120$ K. Three different values for the exchange parameter $J$ are shown. The best fit gives $J= 195$ K. }
\end{center}
\end{figure}

Using the standard notation for the hyperfine coupling in the NMR literature, \cite{ImaiLadder,ItohCuGeO3} let $\widetilde{B}_{(\alpha)} = B_{(\alpha)} / g \gamma_{n}\hbar$.
From the measurements of the spin susceptibility \cite{Imai2011} and the Knight shift \cite{FuScience}, we obtain the transferred hyperfine coupling tensor for oxygen to be $(\widetilde{B}_{(a)}, \widetilde{B}_{(a^{*})}, \widetilde{B}_{(c)}) = (33, 43, 36)$~kOe.  
This now leaves the exchange constant $J$ as the only free parameter in our calculations.

The comparison is shown in Fig.~4. We see a very good agreement between experiment and theory
with $J$ in the range $195\pm 20$ K. These values for $J$ are in agreement with other experiments \cite{klhm-e}and with
ab initio calculations \cite{valenti}. The convergence of our calculation breaks down at low temperatures just as the rates show evidence for a sharp drop
with temperature.

{\it Discussions and Conclusions}:
We have used the Numerical Linked Cluster method to calculate the Nuclear Magnetic Relaxation rates in kagome antiferromagnets
in Moriya's Gaussian approximation and beyond.
While the Gaussian approximation gives a qualitatively correct behavior of the temperature dependence of the rates including a spin-gap
like feature at low temperatures,
it underestimates the magnitude of the relaxation rates by up to a factor of two even at temperatures above the exchange energy scale $J$. 
The use of higher moments fixes the discrepency at high temperatures
giving very good quantitative agreement with the experimental data with a value of $J= 195 \pm 20$ K. 
This shows that despite the presence of antisite disorder, the Heisenberg model provides a good quantitative model for intrinsic spin dynamics 
of Kagome planes in Herbertsmithite as obtained in the NMR measurements.

However, the convergence of our calculations with multiple frequency moments clearly breaks down just as the rates appear to decrease
sharply with temperature. Thus, we are unable to theoretically address the existence of a spin-gap in the system.
The breakdown of NLC implies the existence of longer-range dynamic correlations in the system in contrast to very short-range
static correlations. This causes individual moments to not converge within NLC. There is also need for more moments
to capture the more complex frequency dependence expected at low temperatures. We believe that physically this breakdown marks the onset of
a coherent regime, where
the high energy modes, which contribute an increasing amount to the moments, can now propagate coherently
over longer distances. Such a physics is implicit in
the sharp spectral features in the Brillouin Zone seen in neutron scattering experiments \cite{neutron} and the calculations based on Z$_2$ spin
liquids \cite{sachdev}.

One intriguing feature of our calculations is that the rates show a small peak and then a sharp downturn towards zero.
It is seen in Moriya's Gaussian approximation but also obtained
in the extrapolation with multiple moments. Unfortunately, our convergence breaks down around this temperature. Thus, more theoretical
studies, perhaps based on $Z_2$ quantum spin-liquids \cite{sachdev,oleg}, are needed to see if such a behavior is real or an artifact
of these computational methods. 
We note that such a peak is not seen in the experimental data.

We also find that the spectral functions are well described by a Lorentzian lineshape at low frequencies. This implies that the kagome-lattice
Heisenberg model has much larger low frequency spectral weights at finite $T$ than would be expected just from the short time
behavior. When available, it would be interesting to compare our spectral weights with neutron scattering measurements \cite{deVries} in the
temperature range between $50$ and $300$ K, where our calculations are most reliable. Finite temperature DMRG methods, using
minimally entangled thermal states \cite{mets} may allow for the calculation of spectral
functions at still lower temperatures.

In a system like Herbertsmithite, where much of the low temperature and low frequency bulk behavior may be affected by impurities, quantitative
understanding of the {\it local probes such as NMR}  is essential to elucidate the nature of the exciting quantum spin-liquid phase and our calculations represent a
step in that direction.

{\it Acknowledgements}:
This work is supported in part by the US National Science Foundation grant DMR-1306048 (NS and RRPS) and
NSERC and CIFAR in Canada (TI).

\end{document}